# REACTION RATE THEORY WITH ACCOUNT OF THE CRYSTAL ANHARMONICITY

V.I. Dubinko[1], P.A. Selyshchev [2], J. F. R. Archilla[3]

[1]NSC Kharkov Institute of Physics and Technology, Akademicheskaya Str.1, Kharkov 61108, Ukraine
[2] Kiev national Taras Shevchenko university Vladimirskaya Str. 64, Kiev 01033, Ukraine
[3] Seville University, E-41011, Seville, Spain

Reaction rate theory in solids is modified taking into account intrinsic localized modes or discrete breathers (DBs) that can appear in crystals with sufficient anharmonicity resulting in violation of Arrhenius law. Large amplitude oscillations of atoms about their equilibrium positions in the lattice cause local potentials of alternating sign, which are described in terms of time-periodic modulations of the potential barriers for chemical reactions taking place in the vicinity of DBs. The reaction rate averaged over large macroscopic volumes and times including a lot of DBs is increased by a factor that depends on the DB statistics. The breather statistics in thermal equilibrium and in thermal spikes in solids under irradiation with swift particles is considered, and the corresponding reaction rate amplification factors are derived.

## 1  Introduction

The problem of escape from metastable states is of importance to many fields of physics, chemistry, engineering and biology. It is well-known that in thermal equilibrium the fluctuation-activated reaction rate, $\dot{R}$, is expressed by Arrhenius' law:

$$\dot{R} = R_0 \exp\left(- E_a / k_b T\right) \tag{1}$$

where $R_0$ and $E_a$ are the frequency factor and the activation energy, respectively, $k_b$ is the Boltzmann constant and T is the temperature. How can the interplay of nonlinearity and discreteness of the lattice influence this law? It has been shown that in crystals with sufficient anharmonicity a special kind of time-periodic and spatially localized vibrations can appear named intrinsic localized modes (ILMs) or discrete breathers (DBs) [1-5]. DBs have frequencies above or below the phonon band so that they do not couple with phonons and thus are thermally stable. Over the last decade much progress has been achieved in the understanding of DB properties and their role in various experimental situations. MacKay and Aubry [2] suggested that the existence of DBs could result in apparent violation of Arrhenius law, that is, the phenomenon of chemical reactions taking place at much lower temperatures than expected. Further development of this hypothesis by Archilla et al [3] has taken into account the DB statistics [4] for the evaluation of the reaction rate due to the DBs having energies above the activation energy. They have shown that, although there are many less breathers than phonons, there may be many more with energies *above the activation energy*, making them good candidates to explain e.g. low temperature reconstructive transformations observed in some layered insulators. In this paper we show that reaction rates depend on DBs *of all energies* due to effect of the time-periodic modulation of the activation energy. Large amplitude oscillations of atoms about their equilibrium positions in the lattice cause local potentials of alternating sign, which may be described in terms of time-periodic modulations of the potential barriers for chemical reactions taking place in the vicinity of DBs.

The paper is organized as follows. In Section 2 we present a generic model for a Brownian particle escape from a potential well with a barrier height periodically modulated in time. In Section 3 we combine the reaction amplification rate due to one DB with the breather statistics in thermal equilibrium and in thermal spikes in solids under irradiation with swift particles. In Section 4 we discuss possible extensions of the present model and some of the outstanding problems. We summarize in Section 5.

## 2 Escape rate with account of the potential barrier modulation

Consider a heavily damped particle of mass $m$ and viscous friction $\gamma$, moving in a symmetric double-well potential $V(x)$ (see Fig. 1). The particle is subject to fluctuational forces that are, for example, induced by coupling to a heat bath. Such a model is archetypal for investigations in reaction-rate theory [6]. The fluctuational forces cause transitions between the neighboring potential wells with a rate given by the famous Kramers rate:

$$\dot{R}_K = R_0 \exp\left(-\Delta V/D\right), \quad R_0 = \frac{\omega_0 \omega_b}{2\pi\gamma}, \tag{2}$$

with $\omega_0^2 = V''(x_m)/m$ being the squared angular frequency of the potential in the potential minima, and $\omega_b^2 = \left|V''(x_b)/m\right|$ the squared angular frequency at the top of the barrier; $\Delta V$ is the height of the potential barrier separating the two minima, $D$ is the Gaussian white noise strength related to the temperature as $D = k_b T$.

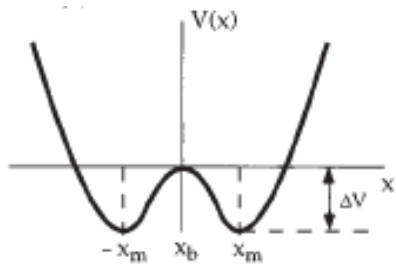

Fig. 1. Sketch of the double-well potential $V(x) = (1/4)bx^4 - (1/2)ax^2$. The minima are located at $\pm x_m$, where $x_m = \left(a/b\right)^{1/2}$. These are separated by a potential barrier with the height given by $\Delta V = a^2/4b$.

In the presence of periodic driving, the double-well potential $V(x,t) = V(x) - V_m(x/x_m)\cos(\Omega t)$ is tilted back and forth, thereby raising and lowering successively the potential barriers of the right and the left well, respectively, in an antisymmetric manner. If the driving frequency is small compared to the thermalization frequency, i.e. if $\Omega << \omega_0$, then periodically modulated escape rates of the Arrhenius type may be used (adiabatic assumption [7]):

$$\dot{R}(t) = \dot{R}_K \exp\left(\frac{V_m \cos(\Omega t)}{D}\right) \tag{3}$$

This expression is widely used in the theory of stochastic resonance [7], which shows that noise-induced hopping between the potential wells can become synchronized with the weak periodic forcing. We are interested in another aspect of this problem, namely, in the rate of escape from the well averaged over the modulation period, $2\pi/\Omega$, which is given by:

$$\left\langle \dot{R}(t) \right\rangle = \dot{R}_K \frac{\Omega}{2\pi} \int_0^{2\pi/\Omega} \exp\left(\frac{V_m \cos(\Omega t)}{D}\right) dt = \dot{R}_K I_0\left(\frac{V_m}{D}\right) = \left|\frac{heat}{bath}\right| = \dot{R}_K I_0\left(\frac{V_m}{k_b T}\right) \tag{4}$$

where the amplification factor, $I_0(x)$, is the zero order modified Bessel function of the first kind.

In order to evaluate the average escape rate in a more general case of an arbitrary modulation frequency we will use another assumption, i.e. that the probability for a particle to escape from a well in each "jump" is given by $\exp(-\tilde{E})$, where $\tilde{E}$ is a random value that fluctuates around its mean value as $\tilde{E} = \langle E \rangle + E_m \cos(\tilde{\varphi})$ with a probability density for $\tilde{\varphi}$ given by $p(\tilde{\varphi}) = 1/2\pi$. Then the integration over $\tilde{\varphi}$ will give for the average escape rate the same expression as in the adiabatic case:

$$\left\langle \dot{R}(\tilde{\varphi}) \right\rangle = \dot{R}_K \frac{1}{2\pi} \int_0^{2\pi} \exp(E_m \cos(\tilde{\varphi})) d\tilde{\varphi} = \dot{R}_K I_0(E_m), \quad E_m = \frac{V_m}{D}. \tag{4*}$$

This assumption is more general than the adiabatic one, since it requires essentially only the independence of the fluctuational force acting on a particle (characterized by the white noise) from the barrier modulation. In both cases, the amplification factor is determined by the modulation to noise ratio, and it does not depend on the modulation frequency or the mean barrier height.

Thus, although the periodic forcing may be too weak to let the particle roll periodically from one potential well into the other one ($V_m < \Delta V$), it can amplify the average reaction rate drastically if the ratio of the modulation amplitude to the noise strength is high enough, as it is demonstrated in Fig. 2.

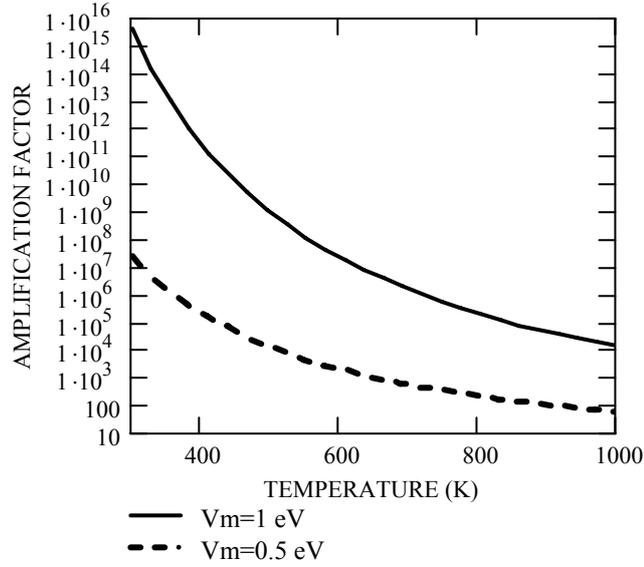

Figure 2. Amplification factor, $I_0(V_m/k_BT)$, for the average escape rate of a thermalized Brownian particle from a modulated potential barrier at different temperatures and modulation amplitudes.

## 3    Breather-induced amplification of reaction rates

Large amplitude oscillations of atoms about their equilibrium positions in the lattice cause local potentials of alternating sign, which may be described in terms of time-periodic modulations of the potential barriers for chemical reactions taking place in the vicinity of DBs. DBs may have frequencies above or below the phonon band depending on the system under consideration. Since the effective breather lifetime is much longer than the atomic oscillation period, we will assume the barrier modulation amplitude to be constant during the breather lifetime and to be proportional to the breather energy. Then the reaction rate at the breather sites will exceed that elsewhere by the amplification factor (4) determined by the breather energy, $E_B$. We are interested in the reaction rate averaged over large macroscopic volumes and times including a lot of DBs. Since DBs can appear at any lattice site randomly, this average can be found by multiplying the concentration (the mean number per site) of breathers of a given energy, $f_B(E)$, into the corresponding amplification factor, $I_0(E)$ and integrating over all possible DB energies normalized to the noise strength:

$$\langle \dot{R} \rangle_B = \dot{R}_K \left( \int_{E_{\min}}^{E_{\mathrm{mod}}} f_B(E) I_0(E) dE + \int_{E_{\mathrm{mod}}}^{E_{\max}} f_B(E) I_0(E_r) dE \right), \qquad E \equiv E_B/D, \qquad (5)$$

$$E_{\mathrm{mod}} = \begin{cases} E_{\max}, & if \ E_{\max} \leq E_r \\ E_r, & if \ E_{\max} > E_r \end{cases}, \qquad (5^*)$$

where $E_r \equiv \Delta V/D$ is the normalized reaction activation energy, $E_{\mathrm{mod}}$ is the maximum barrier modulation amplitude, which is assumed to be equal to the barrier height for breathers with energies higher than that.

In order to evaluate $f_B(E)$ we will use the DB statistics developed in refs [3, 4] for 2D breathers. This statistics theory is based on some simple hypotheses, which, in principle, can be fairly general: (i) DBs in two and three dimensions have a minimum energy $E_{min}$, (ii) The effective lifetime of a breather with energy $E$ is given by $\tau_B(E) = \tau_B^0 \left( \dfrac{E}{E_{min}} - 1 \right)^z$, with $z$ and $\tau_B^0$ being constants (which means that they do not change with the energy $E$) that depend on the system. This law is the simplest mathematical expression that takes into account that large breathers have longer lives than smaller ones, with the lifetime of breathers with minimum energy $E_{min}$ equal to zero. It has to be considered as an approximation because it is not derived from first principles.

In this terms, the rate equation for $f_B(E,t)$ can be written as follows

$$\frac{\partial f_B(E,t)}{\partial t} = K_B(E) - \frac{f_B(E,t)}{\tau_B(E)}, \tag{6}$$

where $K_B(E)$ is the rate of creation of DBs with energy $E > E_{min}$. It has an obvious steady-state solution at $\partial f_B(E,t)/\partial t = 0$:

$$f_B(E) = K_B(E)\tau_B(E), \tag{7}$$

In the following sections we will consider the breather formation by thermal activation and then extend the model to non-equilibrium systems.

### 3.1 DB formation by thermal activation

The rate of creation of DBs is assumed to be proportional to $\exp(-E_B/k_bT)$ since breathers form from fluctuations through an activation process [3, 4]. In normalized energy units one has

$$K_B(E) = K_B^0 \exp(-E), \qquad E \equiv E_B/k_bT, \tag{8}$$

whence it follows that under thermal equilibrium, the DB energy distribution function $f_B(E)$ and the mean number of breathers per site $n_B$ are given by

$$f_B(E) = K_B^0 \tau_B^0 \left( \frac{E}{E_{min}} - 1 \right)^z \exp(-E), \tag{9}$$

$$n_B = \int_{E_{min}}^{E_{max}} f_B(E)dE = K_B^0 \tau_B^0 \frac{\exp(-E_{min})}{(E_{min})^{z+1}} \int_0^{E_{max}-E_{min}} y^z \exp(-y)dy, \tag{10}$$

Noting that $\Gamma(z+1,x) = \int_0^x y^z \exp(-y)dy$ is the second incomplete gamma function [3], eq. (10) can be written as

$$n_B = K_B^0 \tau_B^0 \frac{\exp(-E_{\min})}{(E_{\min})^{z+1}} \Gamma(z+1, E_{\max} - E_{\min}),$$ (11)

It can be seen that the mean DB energy is higher than the averaged energy density (or temperature):

$$\langle E_B \rangle = k_b T \langle E \rangle = k_b T \frac{\int_{E_{\min}}^{E_{\max}} f_B(E) E dE}{\int_{E_{\min}}^{E_{\max}} f_B(E) dE} \xrightarrow{E_{\max} \to \infty} k_b T (E_{\min} + z + 1),$$ (12)

So far we have followed the reasoning of ref [3], in which the mean reaction rate due to DBs is assumed to be determined by breathers with energies higher than the reaction activation energy (the potential barrier height) and hence it can be written as follows:

$$\langle \dot{R} \rangle_B = R_0^B \int_{E_r}^{E_{\max}} f_B(E) dE = R_0^B K_B^0 \tau_B^0 \frac{\exp(-E_{\min})}{(E_{\min})^{z+1}} \int_{E_r - E_{\min}}^{E_{\max} - E_{\min}} y^z \exp(-y) dy,$$ (13)

where $E_r \equiv \Delta V / k_b T$ is the normalized reaction activation energy and $R_0^B$ is the frequency factor that may be different from $R_0$ in eq. (1) and it should be determined separately. It is evident that in this model, the breather effect on reaction rate vanishes if $E_{\max} \leq E_r$.

According to the present model, DBs of all energies influence the reaction rate due to their persistent character resulting in the reaction barrier modulation effect. Combining (9) and (5), the mean reaction rate averaged over all breather energies takes the form:

$$\langle \dot{R} \rangle_B = \dot{R}_K \langle A \rangle_B, \quad E_{\text{mod}} = \begin{cases} E_{\max}, & \text{if } E_{\max} \leq E_r \\ E_r, & \text{if } E_{\max} > E_r \end{cases},$$ (14)

$$\langle A \rangle_B = K_B^0 \tau_B^0 \left( \int_{E_{\min}}^{E_{\text{mod}}} \left( \frac{y}{E_{\min}} - 1 \right)^z \exp(-y) I_0(y) dy + \int_{E_{\text{mod}}}^{E_{\max}} \left( \frac{y}{E_{\min}} - 1 \right)^z \exp(-y) I_0(E_r) dy \right),$$ (15)

where $\langle A \rangle_B$ is the averaged amplification factor.

In order to make quantitative estimates one has to know the product $K_B^0 \tau_B^0$ and other parameters of DBs ($z$, $E_{\min}$, $E_{\max}$) that depend on the system. We will use the results of numerical simulations [3], in which the numbers of breathers and their energy spectra have been calculated for a two-dimensional network of 50 x 50 nonlinear oscillators cooled down from the initial temperature of about 600 K. The mean number of breathers per site $n_B$ was about $10^{-3}$, and according to the energy spectra, all types of multibreathers and single breathers with different symmetries have been formed with the following range of parameters: $z \approx 0.5 \div 3$, $E_{\min} \approx 4 \div 16$, $E_{\max} \approx 10 \div 20$. In order to fit approximately the numerical form of $f_B(E)$, six breather types have been introduced, each one characterized by its own parameters $z$, $E_{\min}$, $E_{\max}$ and a relative probability to occur, $p$ (Table 1). Substituting this data in eq. (11), the product $K_B^0 \tau_B^0$ can be estimated for each DB type and used for the evaluation of the DB amplification factor.

| $E_B^{\min} = E_{\min} k_b T$ (eV) | 0.24 | 0.37 | 0.41 | 0.62 | 0.67 | 0.83 |
|---|---|---|---|---|---|---|
| $z$ | 1.5 | 1.17 | 3 | 0.52 | 2.07 | 1.8 |
| $E_B^{\max} = E_{\max} k_b T$ (eV) | - | 0.47 | - | - | - | 0.95 |
| $p$ | 0.103 | 0.026 | 0.281 | 0.097 | 0.202 | 0.290 |
| $K_B^0 \tau_B^0$ | 0.35 | 2.3 | 773 | $1.2 \times 10^3$ | $1.8 \times 10^5$ | $9 \times 10^6$ |

Table 1. DB parameters deduced from numerical simulations [3].

Fig. 3 shows the amplification factor dependence on the reaction activation energy according to the present model and the model [3] assuming that all DBs have the same maximum energy. The comparison demonstrates that the modulation effect rapidly increases with increasing reaction barrier up to the maximum DB energy, above which it becomes the only mechanism of the reaction rate amplification. In this region the amplification factor does not depend on the activation energy.

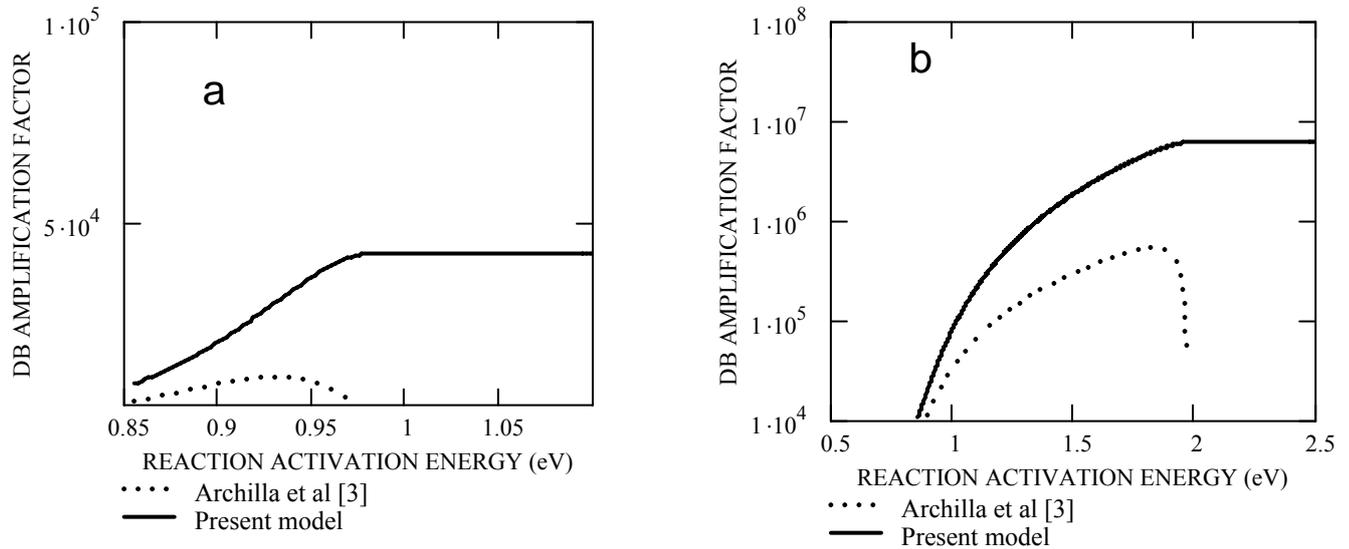

Figure 3. Amplification factor dependence on the reaction activation energy at T = 600 K according to the present model (eq. (15) and the model [3] (eq. (13)) assuming $R_0^B = R_0$ and $E_B^{\max} = 1$ eV (a); $E_B^{\max} = 2$ eV (b) for all breather types. Other DB parameters are presented in Table 1.

In this section as in the refs [3, 4], the rate of breather creation was assumed to be proportional to $\exp\left(-E_B/k_b T\right)$. However, the necessary conditions for breather creation are not yet clear. Some studies suggest that DBs in thermal equilibrium will be generated only if the averaged energy density (or temperature) is large enough for nonlinear terms in the equations of motion to be relevant [8]. In the opposite case of rather small temperatures no breathers are expected to persist. As was argued in ref [4], the basic mechanism leading to energy localization is modulation instability of short wavelength modes. The latter is more effective if dissipation of long wavelength phonons is fast enough, which may explain the principal role of cooling (i.e. transient process of relaxation of a non-equilibrium state towards equilibrium) in the breather formation. A prominent example is the relaxation of an initially strongly heated lattice part modeled in ref [8]. In the following section we will consider a natural example of such processes that take place in solids under irradiation with swift particles. Apart from fundamental aspects, this example is of considerable technological importance in the fields of nuclear engineering and radiation effects.

### 3.2 DB formation in radiation-induced thermal spikes

When irradiating materials by swift heavy ions, the input energy is mainly transferred to the electrons of the target and there is no direct atomic displacement like in the case of irradiation in the nuclear collision regime. Since more than 90% of the energy loss of such projectiles is due to the electron excitations [9] the corresponding structural changes are often related to the so called ``thermal spikes'' (TSs) [10, 11] resulting from energy and momentum transfer from the excited electrons to the lattice in the track. As a result, thermally activated processes are enhanced, which should be taken into account in modeling of the radiation effects.

At the times of about $10^{-15}$ s electron gas within the trek becomes thermolized and its temperature can reach 10-100 eV [10, 11]. This thermal spike dissipates due to the heat diffusion in electron system and due to the heat transfer to the ion system, which is described conventionally by a set of two differential equations of the heat diffusion and exchange [12]:

$$\frac{\partial T_e}{\partial t} = \chi_e \Delta T_e - \alpha_e \left(T_e - T_p\right); \quad \alpha_e \equiv \frac{\alpha_{ep}}{c_e}, \tag{16}$$

$$\frac{\partial T_p}{\partial t} = \chi_p \Delta T_p + \alpha_p \left(T_e - T_p\right), \quad \alpha_p \equiv \frac{\alpha_{ep}}{c_p}, \tag{17}$$

subjected to the initial conditions $T_{e,p}\left(t=o\right)=\theta_{e,p}\left(\vec{r}\right)$, where $c_{e,p}$ and $\chi_{e,p}$ are the heat capacity and conductivity of each subsystem, $\alpha_{ep}$ is the electron-ion coupling coefficient.

Since the time of the energy transfer from a particle to the medium is much less that the relaxation times of the medium, one may consider the particle to be an instant source of energy that causes some initial temperature distribution in electron and phonon systems given by $\theta_{e,p}(\vec{r})$, where the subscript "e" corresponds to electrons and "p" corresponds to phonons. For a sufficiently long trek in Z-direction with axial symmetry of the temperature distribution one has

$$\theta_{e,p}(\vec{r}) = \frac{\varepsilon_{e,p}}{c_{e,p}L}\frac{1}{4\pi\rho_0^2}\exp\left(-\frac{\rho^2}{4\rho_0^2}\right),\tag{18}$$

where $\rho$ is the radius vector modulus, $\rho_0$ is the straggling coefficient, $L$ is the trek length, $\varepsilon_{e,p}$ is the energy transferred to electrons or phonons, respectively.

Assuming the coefficients $\alpha$, $\chi$, $c$ to be temperature independent, an analytical solution of the equation set could be obtained [12, 13]. Consider two limiting cases, which allow making statistical analysis of the effect of thermal spikes on reaction rates that depend exponentially on the lattice temperature.

The limiting case, $\alpha_{ep}\to 0$, describes an adiabatic regime, in which there is no energy transfer between the subsystems, and the solution is given by [12]:

$$T_{e,p}(\rho,t) = \frac{\varepsilon_{e,p}}{4\pi c_{e,p}L(\chi_{e,p}t+\rho_0^2)}\exp\left(-\frac{\rho^2}{4(\chi_{e,p}t+\rho_0^2)}\right),\tag{19}$$

This regime can be realized at short times at which the heat exchange between the subsystems can be neglected:

$$t << t^*, \quad t^* = \left(\alpha_e + \alpha_p\right)^{-1},\tag{20}$$

An opposite (asymptotic) regime, $\alpha_{ep}\to\infty$, can be realized at sufficiently large times and distances from the trek axis:

$$t >> t^*, \quad \rho >> \rho^*, \quad (\rho^*)^2 = \frac{\chi_e+\chi_p}{\alpha_e+\alpha_p};\tag{21}$$

In this regime, electron system cools down to the lattice temperature, and subsequently both systems have a common temperature, $T \approx T_e \approx T_p$, given by the following expression, which is similar to (19) but it has effective heat conductivity, $\chi_a \equiv \left(\chi_e c_e + \chi_p c_p\right)/c_a$, and heat capacity, $c_a \equiv c_e + c_p$:

$$T \approx \frac{1}{4\pi L}\frac{\varepsilon_a}{c_a(\chi_a t+\rho_0^2)}\exp\left(-\frac{\rho^2}{4(\chi_a t+\rho_0^2)}\right), \quad \varepsilon_a \equiv \varepsilon_e + \varepsilon_p,\tag{22}$$

Now, we are interested in the reaction rate, $\langle\dot{R}\rangle_T$, averaged over large macroscopic volumes and times $dVdt$ including a lot of TSs. A mathematical scheme of this averaging procedure has been developed by Lifshits et al [12]. It reduces the many-body problem of TSs in the phase volume $dVdt$ to the one-body problem, i.e. to the temperature field from one spike. The main point of this scheme is the determination of a 4-dimensional volume in space and time, $\Omega(T)$, in which temperature is higher than $T$. In the case of cylindrical spike of a length $L$, $\Omega(T)$ is given by [12]:

$$\Omega(T) = \pi L \int_0^{t_0} \rho^2(T,t)dt \,, \tag{23}$$

where $\rho(T,t)$ is the curve of constant temperature and $t_0(T)$ is the time at which $\rho(T,t)$ falls down to zero. Then the mean reaction rate can be written in the following form:

$$\left\langle \dot{R}(T) \right\rangle_T = \int_0^\infty \dot{R}(\bar{T} + T)P(T)dT \,, \quad P(T) = -\frac{dW(T)}{dT}, \quad W(T) \equiv K_T \Omega(T) = \int_T^\infty P(\theta)d\theta \,, \tag{24}$$

where $W(T)$ is the probability of the temperature deviation from the mean value, $\bar{T}$, by more than $T$; $K_T = J/L$ is the TS production rate, which is proportional to the flux of energetic particles, $J$. The expression for $\left\langle \dot{R}(T) \right\rangle_T$ can be rewritten as the integral over $W$:

$$\left\langle \dot{R}(T) \right\rangle_T = \int_0^1 \dot{R}(\bar{T} + T(W))dW \,, \tag{25}$$

Thus, in order to evaluate $\left\langle \dot{R}(T) \right\rangle_T$ one needs to derive the dependence $\rho(T,t)$ from the solution $T(\rho,t)$ given by (19) and (22), then solve the equation $\rho(T,t_0) = 0$ with respect to $t_0$ and find expressions for $W(T)$ and $T(W)$. In the adiabatic regime this procedure results in the following expressions for $W(T_{e,p})$ in each isolated sub-system [12]:

$$W(T_{e,p}) \approx \left(\frac{T_{e,p}^*}{T_{e,p}}\right)^2, \quad \left(T_{e,p}^*\right)^2 \equiv \frac{1}{16\pi}\left(\frac{\varepsilon_{e,p}}{L}\right)^2 \frac{J}{\chi_{e,p}\left(c_{e,p}\right)^2} \,, \tag{26}$$

In the asymptotic regime one has a similar expression for a common temperature probability distribution [13]:

$$W(T) \approx \left(\frac{T_a^*}{T}\right)^2, \quad \left(T_a^*\right)^2 \equiv \frac{1}{16\pi}\left(\frac{\varepsilon_a}{L}\right)^2 \frac{J}{\chi_a c_a^2} \,, \tag{27}$$

Consider this regime in more details, since it describes the effect of thermal spikes in both subsystems on the reaction rate that depends on the lattice temperature. The substitution of (1) and (27) into (25) results in the following expression for the mean reaction rate:

$$\left\langle \dot{R}(T) \right\rangle_T \approx \left(1 - W(\bar{T})\right)\dot{R}(\bar{T}) + 2R_0\left(\frac{k_b T_a^*}{E_a}\right)^2 \Gamma\left(2, \frac{E_a}{k_b\bar{T}}\right), \quad \Gamma(2,x) = \int_0^x y\exp(-y)dy \tag{28}$$

The probability of the temperature deviation from the mean value is usually very small ($W(\overline{T}) \ll 1$), and so the first term in (28) corresponds simply to the Arrhenius' low, which depends exponentially on the mean temperature. The second term in (28) describes the TS-induced addition to the mean reaction rate, $\left\langle \dot{R}(T) \right\rangle_{TS}$. For sufficiently high activation energy ($E_a/k\overline{T} \gg 1$) it takes especially simple form:

$$\left\langle \dot{R}(T) \right\rangle_{TS} \approx 2R_0 \left( \frac{k_b T_a^*}{E_a} \right)^2 \Gamma\left(2, \frac{E_a}{k_b \overline{T}}\right) \xrightarrow[\frac{E_a}{k_b \overline{T}} \gg 1]{} 2R_0 \left( \frac{k_b T_a^*}{E_a} \right)^2 , \tag{29}$$

which is proportional to the irradiation flux, $J$, and does not depend on the lattice temperature. It is also proportional to the sum of the energies transferred to both subsystems squared, $\left( \varepsilon_e + \varepsilon_p \right)^2$. So even at $\varepsilon_p \ll \varepsilon_e$, when the initial energy transfer to the lattice is negligible, the reaction rate is increased due to the energy transfer via the electron subsystem. The TS effect on the reaction rate (1) is shown in Fig. 4 for the case of a typical ion irradiation.

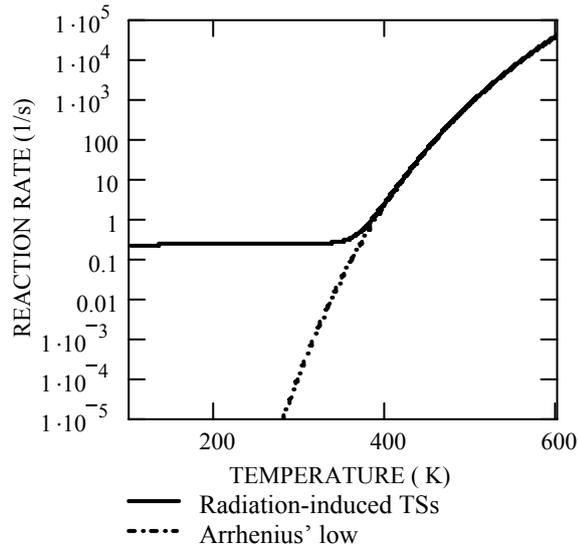

Figure 4. Dependence of the mean reaction rate on the mean lattice temperature given by (28) for the case of ion irradiation of a metal target at the following irradiation and material parameters: $J = 10^{14}$ ion/cm$^2 s$, $\varepsilon_e + \varepsilon_p = 1$ MeV, $L = 2 \times 10^{-4} cm$, $A = 10^{13}$ s$^{-1}$, $E_a = 1$ eV.

It can be seen that the mean reaction rate becomes temperature independent below some threshold temperature, $T_{TS}$, which increases with increasing activation energy and irradiation flux as

$$T_{TS} \approx \frac{E_a}{k_b \ln\left\{ \frac{1}{2} \left( \frac{E_a}{k_B T_a^*} \right)^2 \right\}} \tag{30}$$

This fluctuating temperature background favors the DB creation in the cooling phase of radiation-induced thermal spikes. Averaging the local rate of creation of DBs (8) over large macroscopic volumes and times including a lot of TSs one obtains the expression for the mean production rate of DBs with energy (normalized to the mean temperature) $E = E_B / k_b \overline{T}$ :

$$\langle K_B \rangle_{TS} (E) \approx 2 K_B^0 \left( \frac{T_a^*}{\overline{T}} \right)^2 E^{-2}, \qquad (31)$$

Note that it is inversely proportional to the DB energy squared in contrast to the exponentially strong dependence of the local rate of creation of DBs. It follows from (31) (with account of (7)) that the energy distribution function and the mean steady-state concentration of breathers created in thermal spikes are given by

$$\langle f_B \rangle_{TS} (E) = 2 K_B^0 \tau_B^0 \left( \frac{T_a^*}{\overline{T}} \right)^2 \left( \frac{E}{E_{\min}} - 1 \right)^z E^{-2}, \qquad (32)$$

$$\langle n_B \rangle_{TS} = 2 K_B^0 \tau_B^0 \left( \frac{k_b T_a^*}{E_B^{\min}} \right)^2 \int_1^{E_{\max}/E_{\min}} y^{-2} (y - 1)^z \, dy, \qquad (33)$$

Temperature dependence of the mean concentration of DBs formed by thermal spikes (33) and by thermal fluctuations (9) is shown in Fig. 5. For a sake of simplicity, only one type of DB was selected with parameters that would correspond to the mean DB concentration of $\sim 10^{-3}$ with a mean DB energy of $\sim 0.7$ eV found in the numerical simulations [3] at $T \approx 550 \ K$ .

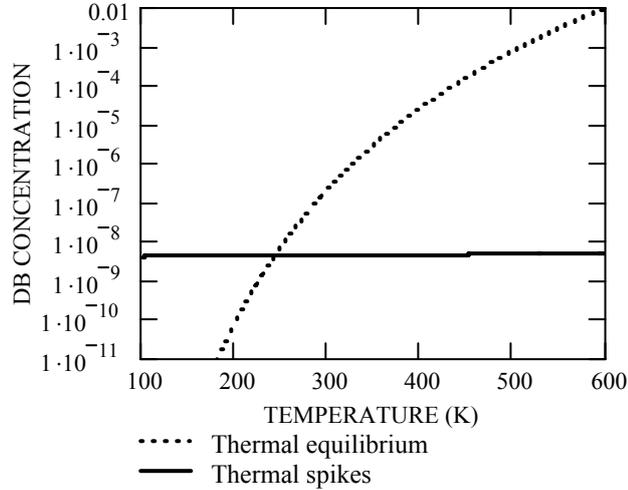

Figure 5. Temperature dependence of the mean concentration of DBs formed by thermal fluctuations (9) and by thermal spikes (33) at the following irradiation and DB parameters: $J = 10^{14}$ ion/cm$^2 s$ , $\varepsilon_e + \varepsilon_p = 1$ MeV , $L = 2 \times 10^{-4} cm$ , $K_B^0 \tau_B^0 = 1$ , $E_B^{\min} = 0.2$ eV , $E_B^{\max} = 1$ eV , $z = 9$ .

Combining (32) and (5), the mean reaction rate averaged over all DBs created in TSs takes the form:

$$\langle \dot{R}(T) \rangle_{B,TS} = \dot{R}_K \int_{E_{\min}}^{E_{\max}} \langle f_B \rangle_{TS} (E) I_0 (E) dE = \langle A \rangle_{B,TS} \dot{R}_K, \qquad (34)$$

$$\left\langle A\right\rangle_{B,TS} = 2K_B^0\tau_B^0\left(\frac{k_bT_a^*}{E_B^{\min}}\right)^2\left(\int\limits_1^{E_{mod}/E_{\min}} y^{-2}\left(y-1\right)^z I_0\left(yE_{\min}\right)dy + \int\limits_{E_{mod}/E_{\min}}^{E_{\max}/E_{\min}} y^{-2}\left(y-1\right)^z I_0\left(E_r\right)dy\right),$$

$$(35)$$

Temperature dependence of the mean reaction rates with account of DBs formed by thermal fluctuations (14) and by thermal spikes (34) is shown in Fig. 6 along with the mean reaction rate due to TS without account of DBs (29). It can be seen that in spite of the fairly low concentration of DBs formed by thermal spikes, their contribution to the TS-induced enhancement of the reaction rates can be very significant.

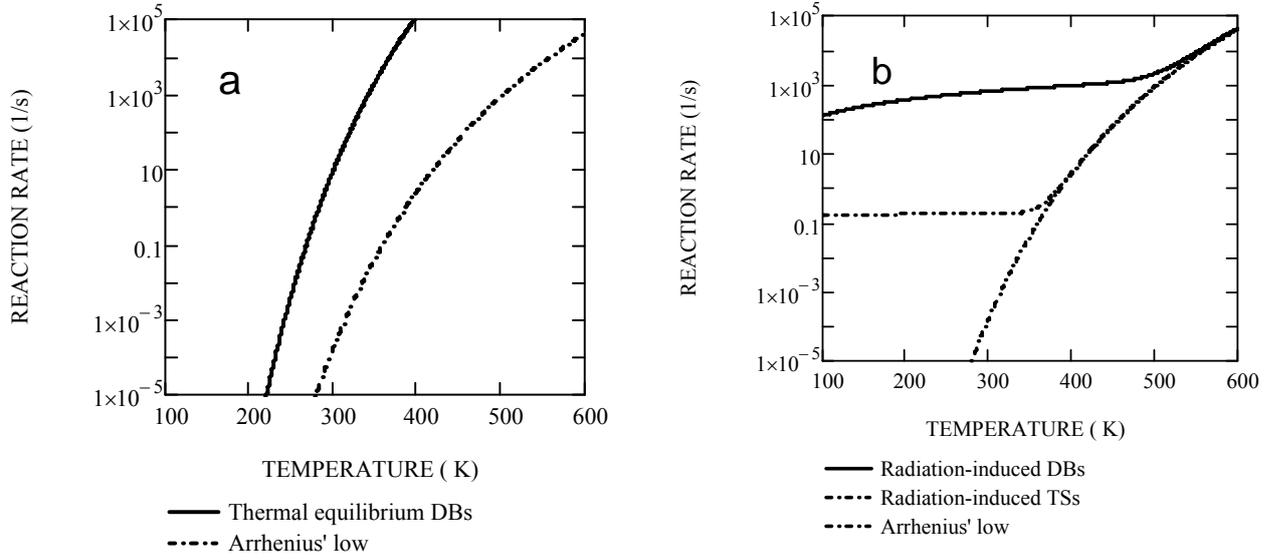

Figure 6. Temperature dependence of the mean reaction rates with account of DBs formed at thermal equilibrium (a) and in thermal spikes (b) at the following irradiation and DB parameters: $J = 10^{14}$ ion/cm$^2 s$, $\varepsilon_e + \varepsilon_p = 1$ MeV, $L = 2\times10^{-4} cm$, $K_B^0\tau_B^0 = 1$, $E_B^{\min} = 0.2$ eV, $E_B^{\max} = E_a = 1$ eV, $z = 9$.

Naturally, the DB effect depends on their parameters ($K_B^0\tau_B^0$, $z$, $E_{\min}$, $E_{\max}$) that depend on the system. To show some of the trends let us express the eq. (34) in the following form

$$\left\langle\dot{R}(T)\right\rangle_{B,TS} = \left\langle A\right\rangle_B\left\langle\dot{R}(T)\right\rangle_{TS},$$ $$(36)$$

$$\left\langle A\right\rangle_B = K_B^0\tau_B^0\left(\frac{E_a}{E_B^{\min}}\right)^2\exp(-\frac{E_a}{k_b\bar{T}})\left(\int\limits_1^{E_{mod}/E_{\min}} y^{-2}\left(y-1\right)^z I_0\left(yE_{\min}\right)dy + \int\limits_{E_{mod}/E_{\min}}^{E_{\max}/E_{\min}} y^{-2}\left(y-1\right)^z I_0\left(E_r\right)dy\right),$$

$$(37)$$

where the amplification factor $\left\langle A\right\rangle_B$ does not depend on irradiation conditions and is determined only by DB parameters and by the reaction activation energy, similar to that for thermal equilibrium breathers defined by eq. (15). Their comparison is shown in Fig. 7.

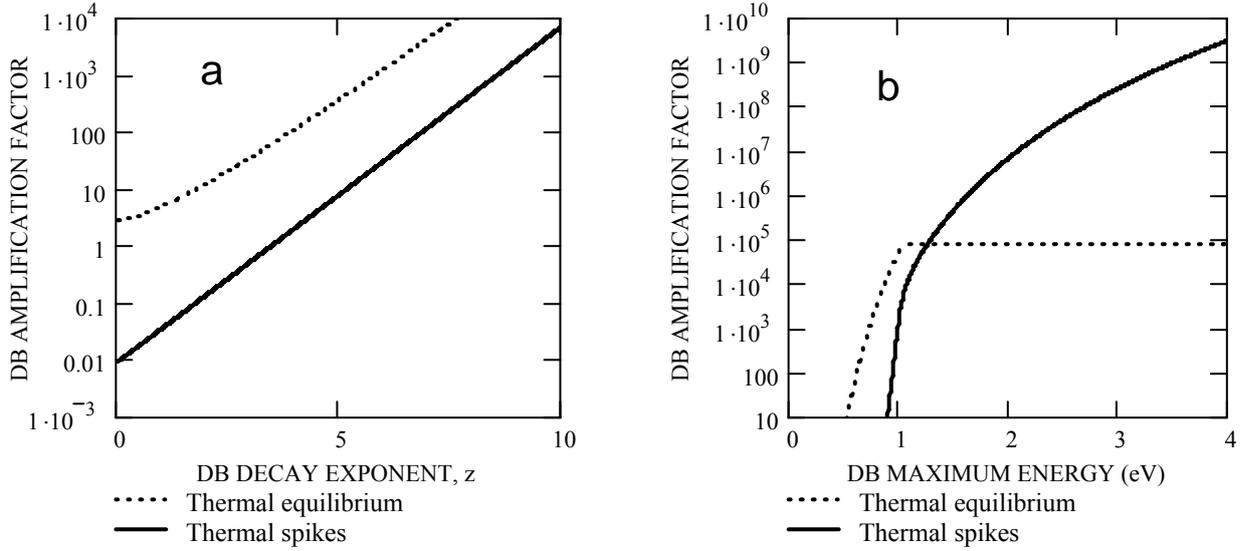

Figure 7. Dependence of the DB amplification factors at thermal equilibrium (15) and under irradiation (37) at room temperature on the DB decay exponent and maximum energy at the following parameters: $K_B^0 \tau_B^0 = 1$, $E_B^{min} = 0.2$ eV, $E_a = 1$ eV; (a) $E_B^{max} = E_a = 1$ eV; (b) $z = 9$.

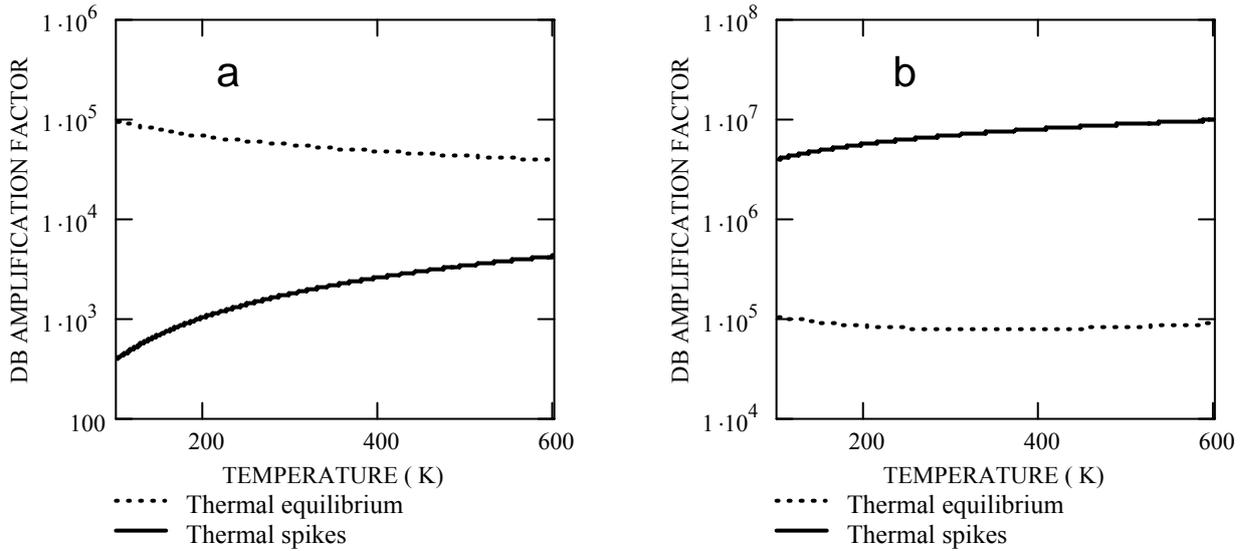

Figure 8. Temperature dependence of the DB amplification factors at thermal equilibrium (15) and under irradiation (37) at the following parameters: $K_B^0 \tau_B^0 = 1$, $E_B^{min} = 0.2$ eV, $z = 9$, $E_a = 1$ eV; (a) $E_B^{max} = 1$ eV; (b) $E_B^{max} = 2$ eV.

The common trend is a very strong dependence on the DB decay exponent, whereas the dependence on the maximum DB energy is qualitatively different. The concentration of high energy breathers at thermal equilibrium (if any) is exponentially low, and so their contribution to the reaction rate is insignificant. On the other hand, the DB production rate in thermal spikes (31) is inversely proportional to the DB energy squared, and so high energy breathers play a major role in the reaction rate amplification due to their relatively longer life times. Temperature dependence of the DB amplification factors is rather weak as demonstrated in Fig. 8 for two different values of the maximum DB energy.

## 4    Discussion

In this section we will discuss some perspective extensions of the present model.

### 4.1    Time evolution of breathers

The equation (6) is the simplest master equation for the DB distribution function, which implies that DBs are formed at a rate $K_B(E)$ and do not change for a time $\tau_B(E)$, after which they disappear. However, a breather can change its energy due to the interaction with phonons, electrons other breathers or stable lattice defects. This could be taken into account in the master equation of the following form

$$\frac{\partial f_B(E,t)}{\partial t} = K_B(E) - \frac{\partial}{\partial E}\left[ f_B(E,t)\frac{dE}{dt} \right] - S_B(E),$$    (38)

where $dE/dt$ is the adiabatically slow rate of breather energy change (e.g. due to phonons), and $S_B(E)$ is the DB sink in the energy space due to various collision events. This equation is similar to that used in the theory of nucleation and evolution of the new phase particles in nonequilibrium environment (see e.g. [14]). Evaluation of the master equation (38) is an outstanding problem of a more comprehensive theory.

### 4.2    Moving breathers

The interaction of moving DBs with defects is presently a subject of great interest and can be connected with a number of phenomena observed in crystals under irradiation, such as the radiation-induced migration of point defects in the crystal bulk [15, 16] or the point defect creation at the extended crystal defects [17-19]. Thermally activated DBs considered in the present paper may move randomly from site to site, which would increase a probability of coupling between DBs and defects. This could be taken into account in the master equation (39).

Another type of mobile localized excitations (called quodons), are created in atomic collisions under irradiation with swift particles. As the incident recoil energy is dispersed, on-site potentials and long range co-operative interactions between atoms can result in the creation of quodons, which are high energy mobile longitudinal optical mode DBs that can propagate great distances in atomic-chain directions [17]. The interaction of quodons with extended lattice defects was suggested to result in the radiation-induced recovery processes such as the void shrinkage [18] and self-organization of the void lattice [19]. However the underlying mechanism, i.e. the quodon-induced vacancy emission from voids, needs further investigations that should take into account the present results. This will be done elsewhere.

## 5 Summary

- Reaction rate theory in solids has been modified taking into account intrinsic localized modes or discrete breathers that can appear in crystals with sufficient anharmonicity resulting in violation of Arrhenius law.

- The reaction rate averaged over large macroscopic volumes and times including a lot of DBs can be increased by many orders of magnitude depending on the DB statistics.

- The breather statistics in thermal equilibrium and under irradiation with swift particles has been considered, and the corresponding reaction rate amplification factors have been derived.

- The reaction rate amplification factor increases strongly with increasing DB lifetime and the maximum DB energy that depend on the system.

- The present model shows that the effects due to the crystal anharmonicity are of both fundamental significance and of considerable technological importance in the fields of nuclear engineering and radiation effects.

**Acknowledgements**

We thank M. Russell, C. Eilbeck and S. Flach for interesting discussions. This work was supported in part (V.D.) by the STCU-NASU grant #4962. JFRA acknowledge financial support from the MICINN project FIS2008-04848